\documentclass{iaus}
\usepackage{graphicx}

\title[]
{`Pure' Supernovae and Accelerated Expansion of the Universe}

\author[M. V. Pruzhinskaya, E. S. Gorbovskoy \& V. M. Lipunov]  
{M. V. Pruzhinskaya$^1$, E. S. Gorbovskoy$^1$
\and V. M. Lipunov$^1$}

\affiliation{$^1$Sternberg Astronomical Institute, Lomonosov Moskow State University, \\ Universitetskii pr. 13, Moscow, 119992 Russia
 \\ email: {\tt pruzhinskaya@gmail.com}}

\pubyear{2011}
\volume{37} 
\pagerange{}

\begin{document}

\maketitle

\begin{abstract}
A special class of type Ia supernovae that is not subject to ordinary and additional intragalactic gray absorption and chemical evolution has been identified. Analysis of the Hubble diagrams constructed for these supernovae confirms the accelerated expansion of the Universe irrespective of the chemical evolution and possible gray absorption in galaxies. 

\end{abstract}

\firstsection 

\section{Introduction}

A remarkable feature of type Ia supernovae (SNe Ia) is the universality of their light curves and a constant absolute magnitude at maximum. This is explained by the similarity of the physical processes that lead to the outburst phenomenon. Generally, this is the thermonuclear explosion of a C-O white dwarf whose mass has become larger than the Chandrasekhar one as a result of accretion (Schatzman’s mechanism) (\cite{1}) or the merger of two white dwarfs with a total mass larger than the stability limit (\cite{2,3}). Unfortunately, because of the differences in outburst mechanisms and differences in chemical composition and masses of the SN Ia progenitor stars, the observed light curves do differ between themselves. With the appearance of a large number of well studied supernovae, it has emerged that the absolute magnitude at maximum can change within $1^m$. Nevertheless, there are methods that allow the absolute magnitude of each SN Ia at maximum to be determined.

Back in the 1970s, Yu.P. Pskovskii established the existence of a relationship between the decline rate of the light curve and the absolute magnitude for SNe Ia (\cite{4,5}). This relationship is refined continuously as the amount of observational data increases and as the observing facilities improve. The most recent estimations give the absolute magnitudes from the brightness decline rate to within hundredths of a magnitude.

However, despite the present-day light-curve standardization methods, there are doubts in the validity of the `standard-candle' hypothesis. First,
the so-called gray dust, whose absorption is wavelength independent and essentially cannot be taken into account\\ (\cite{6}), could
lead to the dimming of distant supernovae. This can be large dust particles with typical sizes greater than $0.01 \mu{m}$ (\cite{7}). In particular, the amount of such dust is proportional to the star formation rate, which increases into the past, and could produce the apparent fall in the power of distant supernovae. To explain the observed dimming of distant SNe Ia, \cite{7,8} invoked the intergalactic gray dust absorption mechanism. However, ~\cite{9} ruled out this possibility. Furthermore, observations of distant quasars show that even if gray dust is present in intergalactic space, it cannot give absorption greater than $0.1^m$ (\cite{10}).

Gray dust inside host galaxies is a different matter. Its amount can also evolve with age, producing the apparent dimming of supernovae. Such dust is
definitely present in galaxies (\cite{11}).

In addition, the further we look deep into the Universe, the chemically earlier population of stars we see. This is attributable to gradual chemical evolution of the Universe due to the thermonuclear burning of primordial hydrogen and helium into heavier elements in stars. It is quite possible that the SN Ia explosion can depend on the chemical composition of a mainsequence star.

The recent discovery of ultraluminous SNe Ia (see~\cite{12}) confirms the existing spread in supernova luminosities at maximum. The existence of such objects is quite predictable within the model of merging white dwarfs the sum of whose masses is not constant and changes slowly with the Hubble time of the Universe. The point is that the mergers of, on average, more massive white dwarfs than those at present occurred at the early evolutionary stages of the Universe. According to Tutukov's calculations (\cite{6}), the mean energy of SNe Ia must grow with z$>$2 and increase significantly at z$>$8. However, very distant supernovae have not yet been discovered in sufficient quantities to be able to reach conclusions about the effect of the total mass of progenitor stars on the absolute magnitude of supernovae, while this effect is insignificant
for nearby supernovae up to z=1 discovered in large quantities.

To get rid of the possible influence of three factors: gray intragalactic dust, chemical evolution, and the difference in white dwarf explosion mechanisms, we suggest using a special class of supernovae — `pure' supernovae.

\section{THE METHOD FOR SELECTING SNe Ia FREE FROM GRAY DUST ABSORPTION}

The idea of our approach is to use only those supernovae that are at great distances from the host galaxy center. First, the oldest, metal-poor stars
with an age comparable to that of the Universe lie at great distances from the nucleus (or high above the plane if we are dealing with an edge on spiral host galaxy). This automatically leads to a more homogeneous chemical composition of the progenitor stars. Second, SNe Ia far from the galaxy center most likely have a common explosion mechanism, namely the merger of white dwarfs. This is because there are no intermediate-mass stars in the galaxy haloes that could provide the accumulation of matter by white dwarfs in binary systems. Recall that the so-called Schatzman (or SD) mechanism (\cite{1}) suggests the accumulation of mass by a white dwarf up to the Chandrasekhar limit in binary systems with a transfer rate of more than $10^{-8}-10^{-7} M_\odot$/yr. In contrast, the white dwarf merger mechanism in elliptical galaxies provides up to 99\% of SN Ia explosions (\cite{13}). Third, there is no dust in the galaxy haloes. For example, the thickness of the dust layer in our Galaxy does not exceed several kpc even at the edge (15-20 kpc). Of course, in elliptical galaxies dust is absent even deep inside the galaxy; besides, the age and, consequently, chemical composition of elliptical galaxies corresponds well to the first generation metal-poor stars. However, the point is that it is very difficult to determine the type of host galaxies for distant supernovae with a redshift of 1. Therefore, we considered only the supernovae far from the host galaxy center.

By now, detailed photometric and spectroscopic observations of a large number of SNe Ia have been performed. For our work, we used the supernovae from \cite{14,15}. \cite{15} used a sample of 414 supernovae discovered by small old surveys and large ones, including SNLS (SuperNova Legacy Survey), ESSENCE (A Supernova Survey Optimized to Constrain the Equation of State of the Cosmic Dark Energy), and SDSS (Sloan Digital Sky Survey). The light curves were processed by the SALT method based on a synthetic SN Ia spectrum (\cite{16}). The supernovae that did not satisfy the processing conditions (an insufficient number of photometric observations, the availability of data only in one band) were excluded from consideration. The final sample included 307 supernovae for which we determined the apparent and absolute B magnitudes, the B-V color at maximum, and the s factor (\cite{17}), a parameter that allow the light curve being investigated to be reduced to a certain model (average) curve by stretching or compressing the time axis, by the SALT method. The supernovae from \cite{14} are also a compilation of data from various surveys, but, nevertheless, they were all processed by the same method.

In the first step, using the HyperLeda (\cite{18}) and SIMBAD databases, from all supernovae we chose those that were far beyond the host
galaxy. The distance from the galaxy center in units of D25 for the galaxy (i.e., the photometric size of the isophote with a magnitude of 25 per square arcmin in the B band) served as a quantitative criterion. If the distance exceeded D25, then the supernova was considered distant. After an examination of the compiled list, we eliminated the questionable cases of spiral and interacting galaxies where it was generally difficult to
draw the boundary of the host galaxy. In addition to the supernovae far from the galaxy center, we selected the supernovae whose host galaxies were confidently classified as elliptical ones. We then examined the objects discovered by the Hubble Space Telescope (\cite{19,20}) at great distances (z$>$0.2). Here, the sample was produced visually. Two halo supernovae, SN2008gy (\cite{21}) and SN2009nr (\cite{22,23}), discovered in the survey of the MASTER robotized network were added to them. Note that SN2008gy and SN2009nr are the first most thoroughly investigated nearby supernovae that are essentially located in intergalactic space. Thus, a list of supernovae ready for the construction of Hubble diagrams appeared (see table \ref{tab1}).

The coordinates from the SIMBAD database were used to draw up he table. Using the HyperLeda database (\cite{18}), the Supernova Catalog of the Sternberg Astronomical Institute (\cite{Ts}), and the Asiago Supernova Catalogue (\cite{Br}), we found the galaxy number according to the PGC (Principal Galaxy Catalog) nomenclature. The sixth column in the table presents the distances, in units of D25, from the supernovae
to the galaxy center. When calculating this distance, we took into account the fact that the galaxy surface brightness falls with distance as $(1+z)^4$ (the Olbers paradox). This effect is negligible for nearby supernovae, but it should be taken into account for distant objects. $R_c$ is the distance between the supernova and the galaxy center in kpc. The last column gives the criterion according to which a given supernova was included in the sample.

\begin{figure}[t]
\begin{center}
 \includegraphics[width=3.4in]{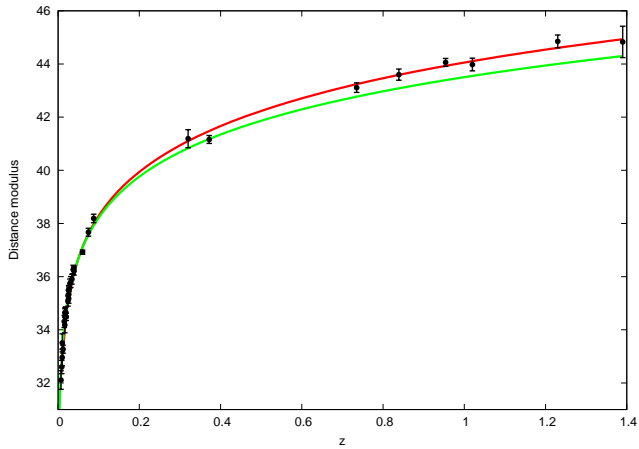} 
 \caption{Hubble diagram for SNe Ia constructed from the points from the table. The red curve is the best fit to the
observational data. The green curve corresponds to the Universe without dark energy.}

   \label{fig2}
\end{center}
\end{figure}

\begin{figure}[]
\begin{center}
\includegraphics[width=3.4in]{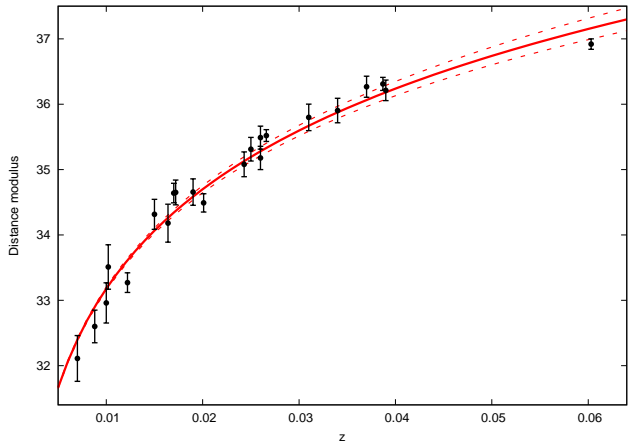} 
 \caption{Hubble diagram for SNe Ia constructed from the points from the table up to z = 0.06.}
   \label{fig1}
\end{center}
\end{figure}

\section{ANALYSIS OF THE HUBBLE DIAGRAMS}

The SNe Ia exploded far from the host galaxy center or in E galaxies were plotted on the Hubble diagram (the redshift dependence of the distance
modulus). To construct the diagram, we used the data from the table. Using the programs written in the MATLAB environment, we were able to fit
the available data by the least-squares method with weights (the $\chi$-squared method). In this case, the difference between the measurement results and their theoretical values was normalized to the standard deviation $\sigma_i$ to take into account the contribution from each individual measurement to the fitting process. The points were best fitted by the following theoretical curve:

\begin{eqnarray}
\mu=\lg(d_l)-5 ~,
\ d_l=(1+z)\frac{c}{H_0}\int\limits_0^z\frac{dz'}{\sqrt{\Omega_m (1+z')^3+\Omega_\Lambda}}
\end{eqnarray}

where $\mu$ is the distance modulus with allowance made for the s factor, the K correction, and absorption; $d_l$ is the photometric distance; $c$, $H_0$ are the speed of light and the Hubble constant. $H_0$ is taken to be 70 km/s/Mpc in the calculations. Since $\Omega_m+\Omega_\Lambda=1$ in the suggested model, the only unknown parameter in the fitting procedure is $\Omega_\Lambda$. The dark energy is chosen in such a way that the sum of the squares of the deviations of the observational points from the theoretical curve (3.1) is minimal. For clarity, we plotted the curve for $\Omega_\Lambda=0$ on the same graph. We see that the slope of the Hubble diagram shows the presence of dark energy, accelerated expansion of the Universe, even for those supernovae that exploded in the regions where the absorption (including the gray one) is minimal! According to these data, the Universe is expanding with $\Omega_\Lambda=0.66\pm0.18$ (see Figure 1).

To answer the question of whether the class of SNe Ia under consideration differs from all of the remaining SNe Ia, we considered the supernovae from
the table in a similar way but up to z$<$0.06 (Figure 2). These supernovae exploded at great distances from the galaxy center or in elliptical galaxies. In both cases, the explosion mechanism is the same: the merger of two white dwarfs. The white dwarf merger mechanism becomes more efficient than the accretion one by two orders of magnitude already a billion years after the formation of an elliptical galaxy (\cite{13}), while all stars far from the center of spiral galaxies have evolved long ago and the accretion mechanism cannot take place. In addition, the chemical composition of these supernovae must be approximately the same. Thus, we have a physically more homogeneous class of supernovae.

\begin{figure}[t]
\begin{center}
\includegraphics[width=3.4in]{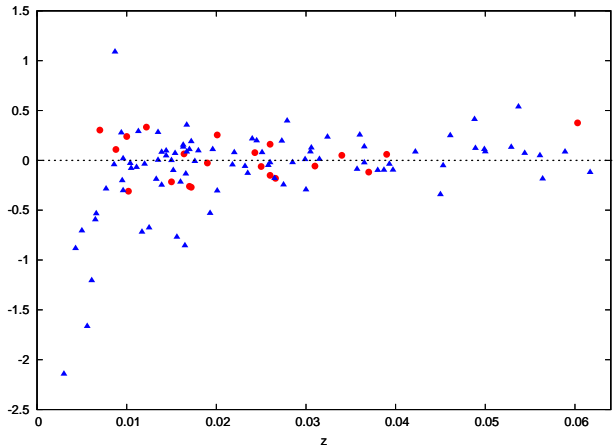} 
\caption{Deviations of the observational points from the model curve on the Hubble diagram. The circles are the `pure' SN Ia and the triangles are the rest SN Ia}
  \label{fig12}
\end{center}
\end{figure}

Figure 3 shows the extent to which the `pure' supernovae and all of the remaining supernovae from \cite{14,15} up to z$<$0.06 deviate from the average curve. The following expression is used for a quantitative estimation of
the variance:

\begin{eqnarray}
\ MSE=\frac{\sum_{i=1}^n (y_i-y)^2}{n-m}
\label{two}.
\end{eqnarray}

where n is the number of supernovae in the sample, m is the number of parameters used in the fitting procedure ($\Omega_\Lambda$), $y_i$ - are the observational points on the Hubble diagram, and y is the theoretical curve fitting these points.

First of all, we note that the variance for `pure' supernovae (z$<$0.06) is considerably lower than that for all supernovae. For example, the mean variance in magnitudes for `pure' supernovae is 0.04, while it is 0.2 for the remaining supernovae from \cite{14,15}.

\section{DISCUSSION AND CONCLUSIONS}

SNe Ia play a great role in various areas of astrophysics. Studying them is very important for the problems of cosmology, because these objects have
turned out to be excellent distance indicators in the Universe in view of their high luminosities and a remarkable similarity of their light curves. They also have shed light on the understanding of the chemical evolution of galaxies by explaining the presence of heavy elements in interstellar space. Nevertheless, the questions related to understanding the explosion physics and the nature of the processes that lead to the supernova phenomenon remain.

As regards the `standard' nature of SNe Ia, here more and more questions arise with every year. Several SN Ia explosion mechanisms have been found
to exist. The brightness of supernovae can change depending on the mechanism being realized. It is also quite possible that the SN Ia explosion can depend on the chemical composition of the progenitor star.

Here, we considered the question of whether the brightness becomes nonstandard due to the absorption of light by gray dust in the host galaxies. According to the gray-absorption hypothesis, the accelerated darkening of supernovae is attributable primarily to the absorption of light by gray dust whose influence is not yet taken into account by the standard procedures of allowance for the absorption (\cite{23}). In addition, the amount of gray dust was greater in the past. To investigate this effect, we considered the supernovae exploded far from the host galaxy center where there is little gas and dust, including gray one. The result obtained shows accelerated expansion of the Universe and the presence of dark energy at a level of $\Omega_\Lambda=0.66\pm0.18$. The difference between the curve corresponding to the Universe without dark energy and the curve fitting the `pure' supernovae on the Hubble diagram is $1^m$ . This value will increase if we take into account the fact that the supernovae were brighter in the past due to the larger mass of the merging white dwarfs.

There is reason to believe that the class of supernovae considered is more homogeneous. These stars exploded in elliptical galaxies or far from the center of their host galaxies. Consequently, the supernova progenitors had a similar chemical composition. The mechanism for the explosion of `pure' supernovae is the merger of two white dwarfs. Therefore, they show a smoother behavior on the Hubble diagram than all of the remaining SNe Ia.

The works to separate the supernovae by the types of galaxies and by the distance from the host galaxy center have already been performed (see \cite{S1,S2,L}). The difference between the observational points on the Hubble diagram and the corresponding values on the curve that fits best these points was found to depend on the type of host galaxy. The scatter is minimal for the supernovae exploded in early type galaxies and maximal for the supernovae exploded in later type and irregular galaxies. The SNe Ia in late type galaxies are, on average, fainter than those in E/S0 ones. The difference in absolute magnitudes for these supernovae is $0.14\pm0.09^m$ (\cite{S1}). The supernovae in spiral galaxies turned out to be redder.

The light-curve optimization algorithms are constructed using the dependence of light-curve parameters on the absolute magnitude of supernovae found
long ago by Yu.P. Pskovskii. However, at this juncture, the above dependence has been investigated incompletely and the parameters entering into it change from work to work. The light-curve processing procedure becomes more complicated as the volume of data increases. Whereas previously the main parameter describing the light curve was its slope, now the changes in the color indices of supernovae are also taken into account. Therefore, it remains to hope that, in the long run, the entire variety of factors affecting the brightness of SNe Ia will be properly taken into account and, thus, these unique natural objects will retain the right to be called `standard candles'.

However, the main result of our analysis is the confirmation of accelerated expansion of the Universe irrespective of the chemical evolution and possible gray absorption. The class of `pure' supernovae we identified can be an efficient tool for investigating the properties of the accelerated expansion of the Universe.

Translated by V. Astakhov

\begin{table}
  \begin{center}
  \caption{List of SNe IA that entered the final sample}
  \label{tab1}
 {\scriptsize
  \begin{tabular}{|l|c|c|c|c|c|c|c|}\hline 
{\bf Name} & {\bf Equatorial coordinates ($\alpha$,$\delta$)} & {\bf z} & {\bf $\mu$} & {\bf PGC} & {\bf D25} & {\bf $\ R_c$} & {\bf Selection criterion}\\ 
\hline
1992bo &01 21 58.44 ---34 12 43.5 & 0.0172 & 34.65 (0.19)   & 4972& 1.10 &25.7 & D25\\
2000bk &12 33 53.94 ---07 22 42.9 &0.0266  &35.52 (0.09)    &41748 &1.16 &32.9 & D25\\
2003fa &17 44 07.72 +40 52 51.6 &0.039   &36.212 (0.157)  &60771&1.43 &39.2& D25\\
2005ms &08 49 14.34 +36 07 47.9 & 0.026  &35.489 (0.176)  &24788 &0.95 &23.1& D25\\
2008bf &12 04 02.90 +20 14 42.6 &0.026   &35.177 (0.178)  &38146 &0.85 &27.3& E galaxy\\
2001ie &10 16 50.70 +60 16 44.5 &0.031   &35.798 (0.203)  &030027 &0.49 &17.0& E galaxy\\
2002dj &13 13 00.34 ---19 31 08.7 &0.010   &32.960 (0.308)  &45908&0.05 &2.0& E galaxy\\
2002do &19 56 12.88 +40 26 10.8 &0.015   &34.315 (0.230)  &63832 &0.22 &2.6& E galaxy\\
2002he &08 19 58.83 +62 49 13.2 &0.025   &35.311 (0.181)  &23371&0.44 &21.0& E galaxy\\
2006nz &00 56 29.21 ---01 13 36.1 &0.037   &36.267 (0.162)  &73507 &0.20 &3.0& E galaxy\\
2007ci &11 45 45.85 +19 46 13.9 &0.019   &34.656 (0.202)  &36670 &0.17 &4.9& E galaxy\\
2008af &14 59 28.50 +16 39 12.3 &0.034   &35.903 (0.187)  &53552 &0.33 &22.1& E galaxy\\
1992au &00 10 40.48 ---49 56 45.3 &0.0603   &36.92 (0.08)  &471591 &0.86 &25.2& E galaxy\\
1997cn &14 09 57.76 +17 32 32.3 &0.0170   &34.64 (0.15)  &050558 &0.12 &4.4& E galaxy\\
1998bp &17 54 50.74 +18 19 50.5 &0.0102   &33.51 (0.34)  &061091&0.15 &3.1& E galaxy\\
1999gh &09 44 19.75 ---21 16 25.0 &0.0088   &32.60 (0.25)  &027885 &0.18& 9.9& E galaxy\\
2000B &07 05 40.73 +50 35 10.5 &0.0201   &34.49 (0.14)  &020136 &0.23 &9.1& E galaxy\\
2000dk &01 07 23.52 +32 24 23.2 &0.0164   &34.18 (0.29)  &003981 &0.65 &3.3& E galaxy\\
1994m &12 31 08.61 +00 36 19.6 &0.0243   &35.08 (0.19)  &041409 &0.30 &14.7& E galaxy\\
1996x &13 18 01.13 ---26 50 45.3 &0.0070   &32.11 (0.35)  &046330 &0.27& 8.9& E galaxy\\
1992br &01 45 44.83 ---56 05 57.9 &0.0876   &38.19 (0.16)  & - &-& -& E galaxy\\
1994am &02 40 02.06 ---01 37 14.9 &0.3720   &41.16 (0.15)  &- &- &-& E galaxy\\
1997ac &08 24 05.21 +04 11 22.6 &0.3200   &41.19 (0.34)  & - &- &-& E galaxy\\
1900Y &03 37 22.59 ---33 02 34.7 &0.0387   &36.31 (0.10)  & - &-& -& E galaxy\\
1992ae &21 28 17.94 ---61 33 01.4&0.0746   &37.67 (0.15)  & - &-&-& E galaxy\\
2009nr &13 10 58.95 +11 29 29.3 &0.0122   &33.27 (0.15)  &45750   &0.71& -& Located in intergalactic space\\
2008gy &03 10 00.96 +19 13 23.1 &0.029   &35.74 (0.17)  &1584648 &- &12.2& Located in intergalactic space\\
HST04 Sas&12 36 54.125 +62 08 22.21&1.39& 44.83 (0.59)&-&- &-&Visually\\
HST05 Spo&12 37 06.53 +62 15 11.70&0.839&43.60 (0.21)&-&- &-&Visually\\
HST05 Fer&12 36 25.10 +62 15 23.84&1.020& 43.98 (0.24)& -&-& -&Visually\\
HST05 Lan&12 36 56.72 +62 12 53.33&1.230&44.85 (0.24)& -&- &-&Visually\\
2003es&12 36 55.39 +62 13 11.9&0.954& 44.06 (0.15)& -&- &-&Visually\\
2002kd&03 32 22.34 ---27 44 26.9&0.735&43.11 (0.18)& -&- &-&Visually\\
\hline
  \end{tabular}
  }
 \end{center}
\vspace{1mm}

\end{table}

\end{document}